\documentclass[aps,twocolumn,superscriptaddress]{revtex4-1}
\usepackage{amsmath,amssymb}
\usepackage{graphics,graphicx}
\usepackage{dcolumn,bm}
\usepackage{psfrag}
\usepackage{color}
\usepackage{multirow}
\newcommand{\cu}
{\affiliation{Department of Physics, University of Calcutta,
92 Acharya Prafulla Chandra Road, Kolkata 700009, India.}}
\newcommand{\rishra}
{\affiliation{Department of Physics, Bidhan Chandra College, 31 Grand Trunk Rd, East,  Rishra, West Bengal 712248, India}}

\newcommand{\vic}
{\affiliation{Physics department, Victoria Institution College, 78B A.P. C. Road,  Kolkata 70009, India.}}

\begin{document}

\title{
Covid-19 spread: Reproduction of data  and prediction using a SIR model on Euclidean network  
}

\author{Kathakali Biswas}
\vic
\author{Abdul Khaleque}
\rishra
\author{Parongama Sen}%
\cu

\begin{abstract}
We study the data 
for the cumulative  as well as  daily  number of 
cases in the  Covid-19 outbreak 
in China.  The  cumulative data can be fit to an empirical form
obtained from a  Susceptible-Infected-Removed (SIR) model studied 
on an Euclidean network previously.  
Plotting the number of cases against the distance from the epicenter 
for both China and Italy, we find an approximate power law variation with an exponent $\sim 1.85$   showing strongly that the spatial dependence plays a key role,
a factor included in the model.  
We report here that the  SIR model on the Eucledean network can reproduce with a 
high accuracy  the  data 
 for China for given parameter values,  and can also  predict when the epidemic, at least locally, can be expected to be over.


\end{abstract}

\maketitle

The 
 novel 
corona virus (COVID-19), which causes an acute respiratory disease in humans
has already spread to nearly 150 countries and has recently been declared 
as a pandemic by the World Health Organisation
\cite{nature,abnormal}.  
The original epicenter has been  
 identified as the city of Wuhan in mainland China and the virus has 
 later spread to
other countries, affecting most severely  Italy and  Iran.   At  present, the  epicenter is believed to have shifted to Europe.
While the rise in the number of cases in the western world is alarming, 
drastic  precautionary  steps  taken in China, South Korea and Hongkong 
have been successful in containing the virus as of now. 
A clear picture of the  space-time dependence of the spread of the virus 
can be understood best from the data of China for which the numbers are quite large.

  Considerable number of analysis of the available data of the number of cases and deaths  have been  attempted recently, 
and a  few data driven models have also  been
proposed \cite{italy,chinese,newsstand,china2,crowd,delay,china3,deep,scaling,effect,artificial,visual,dynamic,trend,ziff,katha}.  
%
In particular, an exponential growth in time in the early stage is noted in all the countries majorly affected,
however, the number of deaths is seen to follow a power law behaviour \cite{ziff}. 
This could be due to purely medical reasons, the possibility of survival also 
depends on the stage of detection and treatment received.

Since the virus can be contracted only once, an affected 
person either dies or recovers;   it belongs to the class of infectious 
diseases  considered in  the 
 Susceptible-Infected-Removed  (SIR) type of models.  
In this model, the population is divided into three categories, Susceptible, 
Infected and Removed;  the total population (including the deaths) is constant. As the disease propagates,
susceptibles are liable to get infected, the  infected people  either die or 
recover (treated in the same manner); the total infected population over time form  the removed category. 
Usually the densities are considered and denoted by $S$, $I$ and $R$ for the three categories and depend on time $t$, with $I = \frac{dR}{dt}$ and $S+I+R = 1$. 
In such epidemics, $I$, corresponding to newly infected 
density in real situations,  initially shows a slow increase with time that changes into a steep rise before 
it reaches  a maximum value.  Typically, the increasing phase shows an exponential behaviour. After reaching the peak,   it gradually decreases before reaching a zero value when one can declare that the epidemic is over. China is already in this decreasing phase while  most other countries 
are yet to reach the peak value. The cumulative data, $R$, accordingly, shows a saturation once the peak value has been crossed. 

\begin{figure}
\includegraphics[width=5cm]{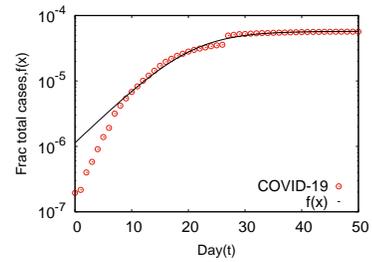}
\caption{The  fraction of total infected people $R(t)$ in China against time.
The fitted curve is of the form given in Eq. (\ref{eq1}).}
\label{china_real}
\end{figure}

In Fig. \ref{china_real}, we plot the data for the total fraction  of cases 
$R$ versus time for China where the  number of total cases is divided by the total population $N_{China}$, 
using the data in \cite{data} where the daily reports are available starting from  
January 21, 2020 (Day 0),  
The number of records  is 50 up to March 11. 
A jump in the data  shows up on the 26th day, this is presumably because  initially, the criteria to be satisfied to confirm that a person 
has actually  contracted the disease had been more stringent  and 
relaxed  later on. 
 
The resultant jump in the data makes it a bit difficult to fit with a smooth function, nevertheless, one can 
 fit the data to  the empirical 
form 
\begin{equation}
R(t)  = a \exp(t/T)/[1+ c \exp(t/T)] 
\label{eq1}
\end{equation}
with $a = 1.15698 \times 10^{-6}$, $c   = 0.0201529$    
and $T   = 5.22 $. 
The error involved for $a$ and $c$ are about 16$\%$ while for $T$ it is $4.3\%$.
%
This was already noted  in an earlier report for the data up to day 40 \cite{katha}. 
The data for the number of newly infected people $I$ has more scatter 
and instead of a direct fitting, we   
calculate it 
by differentiating $R$ with respect to $t$. 
 The peak value of $I$  can be  found out  by calculating the   maximum of $I$ 
and one can also locate the time $t_p$  when the peak occurred: $t_p = T \ln (1/c)$. 
This gives a value 20.38 which is larger than the actual value ($\sim 15$),
the discrepancy   may be due to the discontinuity in the data as mentioned earlier.
Note that the above analysis can be made only when $I$ has entered the decreasing phase.  

\begin{figure}
\includegraphics[width=5cm]{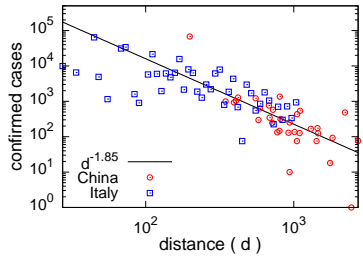}
\caption{The number of cases recorded in a place at a  distance from the epicenter is shown in a log-log plot for China and Italy. Data updated till March 12, 2020.  For Italy, the number of cases is multiplied by 50 for better comparison, these data have also been logarithmically binned.}
\label{fig2:dist}
\end{figure}

The empirical form  given in Eq. (\ref{eq1}) was actually  obtained from a theoretical model of SIR on an 
Euclidean network studied previously \cite{khaleq1} and  was also shown to fit accurately the 
data for the Ebola outbreak in West Africa \cite{khaleq2} that occurred around 5
years back. 
In this model, one assumes that apart from the nearest neighbours, there are also connections randomly at a distance $\ell$ with a probability $P(\ell)$  proportional 
to 
$\ell^{-\delta}$. Here  an infected person can infect her nearest neighbours with a probability $q$.  
Although one assumes that there is no mobility for the agents here,
the fact that they can infect a person at a  distance, implicitly includes the 
possibility  that the agent can travel. 

To justify that one requires a Euclidean model to fit the data, we have also 
calculated the number of cases $R(d)$ recorded as a function of the Haversine distance
$d$, in Km,  from the 
epicenter. 
Taking Wuhan as the epicenter, this data are obtained  for China. 
 Since Italy has also recorded a large number of cases, we  plot the data for Italy alongside, taking Bergamo as the epicenter, where the largest number of cases have been recorded.  
Both show  consistency with a power law decay 
at larger distances as shown in Fig. \ref{fig2:dist};  
\begin{equation}
R(d) \propto d^{-\gamma}
\label{eq2}
\end{equation}
with  $\gamma  = 1.85 \pm 0.1$. Since the data  
understandably have  a large amount of scatter (even after binning), this estimate is approximately 
made. 
While it is expected that $R(d)$ will decrease with $d$, 
 the power law behaviour is   not obvious.
The correlation coefficient is also calculated for the two sets of data; for China it is -0.268  while for Italy the value is -0.383.

This distance dependence, however,  can only be obtained 
with the identification of the epicenter, otherwise 
one obtains no systematic dependence as was the case when the data for the rest of the world were plotted with Wuhan as epicenter \cite{katha}. 

The form of $R$ and the power law dependence of the infection supports the claim that  a Euclidean network model is  appropriate to study the outbreak phenomena.
However, to establish that indeed a  SIR model on a Euclidean network can explain the data, 
we next simulate a ``agent based"  to find the appropriate values, if any, of the parameters that can reproduce the data accurately. For Ebola outbreak,
this was quite successfully done using this simple model.

For the  SIR model on the Euclidean network, we have two parameters, 
$\delta$, already defined,  and $q$, the infection 
probability. The value of $\delta$ essentially denotes the range of contact,  $\delta \geq 2$ corresponds to a short range model.
The model shows long range behaviour for $\delta < 2$ and for sufficiently 
small values of $\delta$ manifests small world behaviour \cite{socio}. For 
real world 
network, $\delta$ is therefore expected to be less than 2, however, the data 
plotted in Fig. \ref{fig2:dist} suggests it is not very far from 2.  
In the model, there exists a   critical value  of the infection probability,  denoted by $q_c$,  above which the disease becomes epidemic. 
$q_c$ obviously depends on $\delta$ and for appropriate reproduction of the 
data, $q$ must be above $q_c(\delta)$.  

\begin{figure}
\includegraphics[width=7cm]{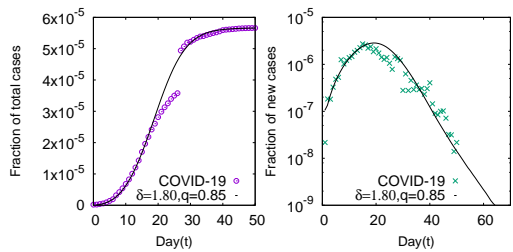}
\caption{Left panel shows the real data for the fraction of population affected versus time (data points) versus the value obtained from the simulation (rescaled). Right panel shows the daily data along with the results from the simulation.}
\label{fig3}
\end{figure}

In the simulation, 100 different network configurations have been used. Initially, one randomly chosen node is infected. For each network, 800 such choices have been considered. A system of  $2^{12}$ nodes has been taken; each node is connected to its nearest neighbours and one more node at a distance randomly on an average.  
To compare the results of the model  with the real data, it is also required to normalise the latter and thus  we
consider the fraction of  total infected people that is plotted in Fig. \ref{china_real}.
 However, this number is rather small even for the saturation value while
in the model it is $\mathcal{O}(0.1)$. Thus one needs to  rescale the data obtained from the model to show the agreement. The time
unit is one Monte Carlo Step (MCS) in the simulation, and also needs  to be rescaled 
according to the real data. 
The rescaling factor for the fraction of affected people 
for the results obtained from the data is determined from the saturation values of the real data and data from the simulations. The rescaling is done such that the two coincide, of course, this does not ensure the entire data will match. 
We also argue that this factor gives an idea of the fraction of the population exposed to the disease. For China, the fraction (rescaling factor)  
$\rho=6.0783 \times 10^{-5}$.  This factor can be interpreted as the fraction of the 
population exposed to the disease.
For the time axis,   we find $T$ in days  $\approx 0.95$ times a Monte Carlo step and the rescaling is done   accordingly. 
 
Fig. \ref{fig3} shows the real data and the results from the model after the rescaling is done 
with the parameters $\delta = 1.8$ and $q =0.85$.  
Here we have also shown the comparison for the new case density. 
We note that the data match quite well with the simulation results, 
apart from the discontinuity in the real data that cannot be reproduced in  the 
simulation.  
Both $\delta$ and $q$ values,  when varied within  $\pm 0.05$,  continue to 
show fairly  good agreement.   

From the model, one can also predict when the disease is going to stop. Ideally this happens when $I$ becomes zero - no  infected person  is left in the population.
Here, if  $I$ becomes less  than $1/N_{China}$, one can assume that
no infected person remains. Accordingly we find that time to be around 64 MCS steps which corresponds to 61 days. 
On the other hand, the simulation shows zero infected agents at 82 MCS steps 
which corresponds to about 78 real days. This indicates  one can expect 
 the disease to be eradicated from China approximately within 60-78 days from January 21, 2020. 
Of course these are averaged values and fluctuations will be there.  

Discussions: We find a rather high value of $\delta$ which gives nice agreement: this may indicate the range of the contact between infected and infectious 
agents is quite short. In principle, $\delta$ can vary from country 
to country depending on density of population, travel patterns and other factors. It is also significant that the
value of $\delta$ is quite close to $\gamma$ (eq. (\ref{eq2})) found in the real data. 

The model, as mentioned earlier, had also reproduced quite well the data for Ebola outbreak. In comparison, we find a larger value of $q$, the infection probability,  here. This is consistent, as in Ebola, the infection is transmitted through more intimate contact.

An important  and interesting point is we obtained good agreement without any
factor representing preventive  measures in the model. Here, how the disease 
is contained in the model should be well understood. In the model, at later stages, the disease gets controlled as an infected person no longer finds enough 
susceptible people to infect. Indirectly this means, the susceptible people
now live in isolation which is exactly what is being aimed at by   imposing curbs on the mobility and social life of the citizens. 
So the model, where mobility is not allowed, in a way successfully mimics this situation. 
Incorporating preventing factors by making e.g, $q$ time dependent or reducing the 
number of neighbours in the model will of course show a flattening of the daily cases curve \cite{amita},  on the other hand one must then also include the possibility 
that the virus mutates and becomes stronger. 
In another recent analysis, the death factor has also been incorporated 
in the SIR \cite{sird}. In any case these factors will make
the model more complicated and our results show it is not necessary.
Also, whether  in reality improvement  in treatment helps in 
controlling the disease is a debatable point. 
 A recent analysis suggests that enhanced medical 
facilities  can  theoretically have an effect in flattening the daily cases
curve,  but not so in practice \cite{noflat}. So social
isolation is the key factor that is ultimately responsible  for the 
control of the disease. 

\medskip

Acknowledgement: 
We thank  Anirban Kundu,  Arnab Chatterjee and Soumyajyoti Biswas 
for interesting discussions and suggestions.
Private  communication with Amita Kapoor and Robert Ziff is also acknowledged.

\end{document}